\documentclass{PoS}
\usepackage{graphicx}

\PoS{PoS(LAT2005)336}

\title{Pion form factors in two-flavor QCD}
\ShortTitle{Pion form factors in two-flavor QCD}

\author{JLQCD collaboration:}

\author{
  \speaker{S.~Hashimoto$^{a,b}$}\thanks{E-mail: shoji.hashimoto@kek.jp},
  S.~Aoki$^c$, 
  M.~Fukugita$^d$, 
  K.-I.~Ishikawa$^e$,
  N.~Ishizuka$^{f,c}$, 
  Y.~Iwasaki$^c$, 
  K.~Kanaya$^c$,
  T.~Kaneko$^{a,b}$, 
  Y.~Kuramashi$^{f,c}$, 
  M.~Okawa$^e$,
  N.~Tsutsui$^a$,
  A.~Ukawa$^{f,c}$, 
  N.~Yamada$^{a,b}$,
  T.~Yoshi\'e$^{f,c}$
  \\
  \llap{$^a$}
  High Energy Accelerator Research Organization (KEK),
  Tsukuba 305-0801, Japan;
  \\
  \llap{$^b$}
  School of High Energy Accelerator Science,
  The Graduate University for Advanced Studies (Sokendai),
  Tsukuba 305-0801, Japan.
  \\
  \llap{$^c$}
  Graduate School of Pure and Applied Sciences,
  University of Tsukuba, Tsukuba 305-8571, Japan.
  \\
  \llap{$^d$}
  Institite for Cosmic Ray Research, University of Tokyo,
  Kashiwa 277-8572, Japan.
  \\
  \llap{$^e$}
  Department of Physics, Hiroshima University,
  Higashi-Hiroshima 739-8526, Japan.
  \\
  \llap{$^f$}
  Center for Computational Sciences, 
  University of Tsukuba, Tsukuba 305-8577, Japan;
}

\FullConference{
  XXIIIrd International Symposium on Lattice Field Theory\\
  25-30 July 2005\\
  Trinity College, Dublin, Ireland}

\abstract{
  We present a calculation of pion electromagnetic and scalar
  form factors in two-flavor QCD with the non-perturbatively
  $O(a)$-improved Wilson fermion. 
  Chiral extrapolation of the corresponding charge radius is
  discussed based on the chiral perturbation theory.
}

\begin{document}

\section{Introduction}
Electromagnetic and scalar form factors of pion are
fundamental quantities in the low energy dynamics of
pions.
In the chiral perturbation theory (ChPT) their radii are
related to the low energy constants at $O(p^4)$.
The electromagnetic charge radius is experimentally measured
rather precisely,
$\langle r^2\rangle_V^\pi$ = 0.452(10)~fm$^2$
\cite{Eidelman:2004wy}, 
and the scalar radius can be related to $F_K/F_\pi$ or
$\pi\pi$ scattering amplitudes.
Therefore, these simple quantities provide a good testing
ground of the lattice calculation techniques.
In particular, the chiral extrapolation seems to induce
large systematic errors unless the unquenched lattice
simulation can treat pions well below 300~MeV where the
chiral logarithm effect should become prominent
\cite{Hashimoto:2002vi,Bernard:2002yk}.
It is interesting to check if the lattice data
reproduce the expected chiral logarithms in the pion form
factors.

Our numerical calculation is performed on the two-flavor
gauge ensembles produced by the JLQCD collaboration
\cite{Aoki:2002uc}
at $\beta$ = 5.2 on a 20$^3\times 48$ lattice with the
$O(a)$-improved Wilson fermions.
High statistics is the key for the form factor measurements;
we accumulated 1,200 gauge configurations for each five
sea quark masses.
The pion form factors are obtained as a by-product of our
calculation of the $K_{l3}$ form factors
\cite{Tsutsui_lat05}.
It turned out that the largest effect on the $K_{l3}$ form
factor at zero momentum transfer $f_+^{K\pi}(0)$ comes from
the shift of the zero recoil point to
$t_{max}=(m_K-m_\pi)^2$.
For the evaluation of $f_+^{K\pi}(0)/f_+^{K\pi}(t_{max})$ we
need a precise knowledge of the pion/kaon charge radius,
which is a subject of this paper.

\section{Pion form factors}
The pion electromagnetic form factor $G_\pi(q^2)$ is defined as
\begin{equation}
  \langle\pi(p')|J_{em}^\mu |\pi(p)\rangle =
  G_\pi(q^2) (p+p')^\mu,
\end{equation}
where $q^2=(p-p')^2$ and $J_{em}^\mu$ is the electromagnetic
current. 
Because of the charge conservation, the form factor
$G_\pi(q^2)$ is normalized as $G_\pi(0)=1$.
It is well known that the experimental data in the small
momentum transfer $Q^2=-q^2\lesssim$ 1~GeV$^2$ are well
described by the vector meson dominance hypothesis
$G_\pi(Q^2) = 1/(1+Q^2/m_\rho^2)$.
(For a summary of experimental results, see for example 
\cite{Blok:2002ew}.)

The charge radius $\langle r^2\rangle_V^\pi$ is a slope of
$G_\pi(q^2)$ near the zero 
momentum transfer
\begin{equation}
  G_\pi(q^2) = 1 + \frac{1}{6} \langle r^2\rangle_V^\pi \, q^2
  + \cdots.
\end{equation}
Near the chiral limit, ChPT predicts the chiral logarithm
\cite{Gasser:1984ux} 
\begin{equation}
  \label{eq:charge_radius_chiral_log}
  \langle r^2\rangle_V^\pi = \frac{12 L_9^r}{f^2}
  - \frac{1}{(4\pi f)^2} 
  \left[ \ln\frac{m_\pi^2}{\Lambda^2} + \frac{3}{2} \right].
\end{equation}
Unlike the pion decay constant, for which the chiral
logarithm appears in the form $m_\pi^2\ln m_\pi^2$, 
the chiral limit of the charge radius is divergent, and one
expects a large effect as the physical pion mass is
approached. 
The current experimental value is
$\langle r^2\rangle_V^\pi$ = 0.452(10)~fm$^2$
\cite{Eidelman:2004wy}.

The scalar form factor $G_S(q^2)$ is defined in a similar
manner. 
\begin{equation}
  \label{eq:scalar_form_factor}
  \langle\pi(p')|\bar{q}q|\pi(p)\rangle = G_S(q^2),
\end{equation}
Here the scalar operator is understood as a flavor
non-singlet one.
The scalar radius $\langle r^2\rangle_S^\pi$ is defined for
$G_S(q^2)$ like $\langle r^2\rangle_V^\pi$ for $G_\pi(q^2)$
in (\ref{eq:charge_radius_chiral_log}).
The chiral logarithm is even stronger for the scalar radius
than for the charge radius, {\it i.e.} the coefficient of
the log term is $-15/2(4\pi f)^2$ rather than 
$-1/(4\pi f)^2$ in (\ref{eq:charge_radius_chiral_log}).

Although there is no direct physical process induced by
(\ref{eq:scalar_form_factor}), the scalar radius can be
extracted from the $\pi\pi$ scattering using ChPT,
$\langle r^2\rangle_S^\pi$ = 0.61(4)~fm$^2$
\cite{Colangelo:2001df}.
It can also be related to $f_K/f_\pi$, from which one
obtains 0.65(4)~fm$^2$.
For non-degenerate quark masses, the scalar form factor
corresponds to $f_0^{K\pi}(q^2)$ of the $K_{l3}$ decay.
Its radius $\langle r^2\rangle_S^{K\pi}$ is measured as 
0.235(14)(8)~fm$^2$ \cite{Yushchenko:2003xz} or
0.165(16)~fm$^2$ \cite{Alexopoulos:2004sy} for charged and
neutral kaons, respectively.
It indicates that the flavor symmetry breaking is a large
effect as suggested by the large chiral logarithm.

\section{Lattice calculation}
For the calculation of pion form factors we calculate
three-point functions $C^{P^SJP^S}(p_\pi,t_J)$ with
operators $J$ (= $V_4$ or $S$) inserted at time $t_J$.
Smeared pion interpolating fields $P^S$ are set at time
$t=0$ and $T/2$ ($T$ is the temporal extent of our lattice
$T=48$).
Spatial momentum is inserted at $t=T/2$ and $t_J$, so that
the initial pion propagating from $t=0$ to $t_J$ is at rest
and the final pion between $t_J$ and $T/2$ has a momentum
$p_\pi$.
To extract the form factor we use a double ratio
\begin{equation}
  \label{eq:double_ratio}
  \frac{
    \displaystyle \frac{C^{P^SJP^S}(p_\pi,t_J)}{C^{P^SJP^S}(0,t_J)}
  }{
    \displaystyle \frac{C^{P^SP^L}(p_\pi,t_J)}{C^{P^SP^L}(0,t_J)}
  } 
  \longrightarrow
  \frac{G_J(q^2)}{G_J(0)} \frac{m_\pi+E_\pi(p_\pi)}{2m_\pi},
  \;\;\;\;\;
  0\ll t_J\ll T/2,
\end{equation}
where the two-point function $C^{P^SP^L}(p_\pi,t)$ in the
denominator is constructed with the same smeared operator
$P^S$ and a local operator $P^L$.
In the large enough time separation the numerator gives the
ratio of form factors $G_J(q^2)/G_J(0)$ up to some
kinematical factor, while the denominator becomes identity
because 
$\langle\pi(p)|P^L|0\rangle/\langle\pi(0)|P^L|0\rangle = 1$ 
must be satisfied for Lorentz invariance.
Positive and negative $t_J$s are averaged in order to
increase statistics.

\begin{figure}[tbp]
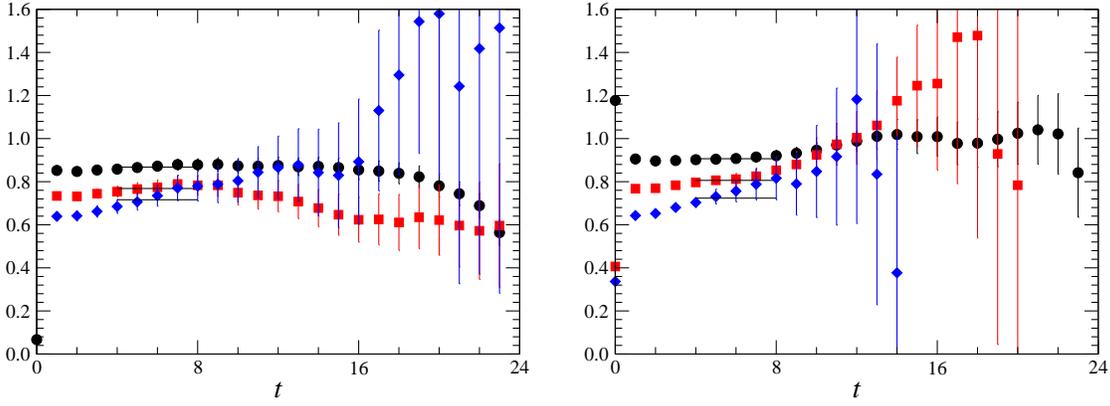

  \centering
  \includegraphics[width=7cm,clip]{figures/V4double_ratio_k1340_11.eps}
  \hspace*{5mm}
  \includegraphics[width=7cm,clip]{figures/V4double_ratio_k1355_55.eps}
  \caption{
    Double ratio (\protect\ref{eq:double_ratio}) for the
    vector form factor at $K$ = 0.1340 (left) and 0.1355
    (right).
    Momentum $p_\pi$ is (1,0,0) (circles), (1,1,0)
    (squares), and (1,1,1) (diamonds) in units of $2\pi/L$.
  }
  \label{fig:double_ratio}
\end{figure}

Plots are shown in Figure~\ref{fig:double_ratio} for the
vector current $J=V_4$.
Results for the heaviest ($K$ = 0.1340) and lightest ($K$ =
0.1355) quark masses are plotted; the data have similar
quality for other quark masses.
(Sea and valence quark masses are set equal in our
calculation setup.)
We find a fairly good plateau beyond $t\simeq 4$, though the
signal for the largest momentum (1,1,1) is not satisfactory.
We fit the data with a constant in the range [4,8] to
extract the form factor ratio in the RHS of
(\ref{eq:double_ratio}).

\section{Form factor results}
\begin{figure}[tbp]
  \centering
  \includegraphics[width=10cm,clip]{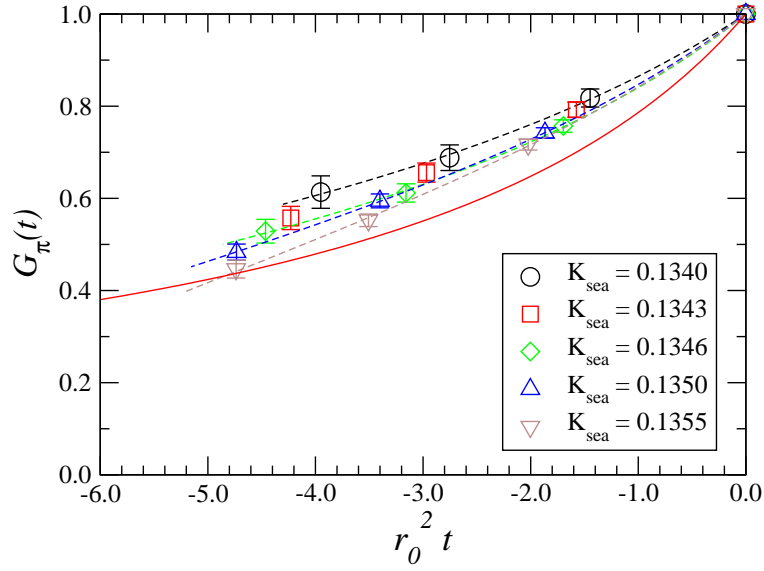}
  \caption{
    Vector form factor as a function of $t=q^2$ normalized
    by the Sommer scale $r_0$.
    Solid curve shows the vector meson dominance form
    $1/(1-q^2/m_\rho^2)$ with the physical $\rho$ meson
    mass.
  }
  \label{fig:vector_form_factor}
\end{figure}

In Figure~\ref{fig:vector_form_factor} we plot the vector
form factor for each $K$ value.
We fit the data with two different fit forms:
\begin{eqnarray}
  \label{eq:free_pole}
  G_\pi(t) & = & \frac{1}{1-c_0 (r_0^2 t)} + c_1 (r_0^2 t)^2,
  \;\;\;\;\;\;\;\;\; \mbox{(free pole)},
  \\
  \label{eq:measured_pole}
  G_\pi(t) & = & \frac{1}{1-t/m_V^2} + d_0 (r_0^2 t),
  \;\;\;\;\;\;\;\;\; \mbox{(measured pole)}.
\end{eqnarray}
Here, $t=q^2$ is normalized by the Sommer scale $r_0$.
$c_0$, $c_1$, and $d_0$ are fit parameters.
Both forms are motivated by the vector meson dominance ansatz,
but the pole mass is a free parameter in the ``free pole''
form (\ref{eq:free_pole}), while the vector meson mass
measured on the lattice at a given quark mass is used in the
``measured pole'' form (\ref{eq:measured_pole}).
In Figure~\ref{fig:vector_form_factor} the fit curves with
the ``free pole'' form are plotted, but both can fit the
lattice data equally well.

Theoretically, we prefer the ``measured pole'' form, because
the vector meson pole must appear with the known vector
meson mass from the analyticity.
There are other contributions from higher resonances and
continuum states, that we may parametrize by the linear term 
$d_0(r_0^2 t)$ near $t=0$.
This statement applies when the valence quark mass is heavy
enough that the $\rho\to\pi\pi$ threshold does not open,
which is the case in our lattice calculation.
As the quark mass decreases, the $\pi\pi$ continuum
state starts to contribute and finally it reproduces the
ChPT prediction near the chiral limit.
Therefore, in the physical quark mass regime the pole
dominance form should be modified appropriately (see, for
example, \cite{Guerrero:1997ku}).

\begin{figure}[tbp]
  \centering
  \includegraphics[width=10cm,clip]{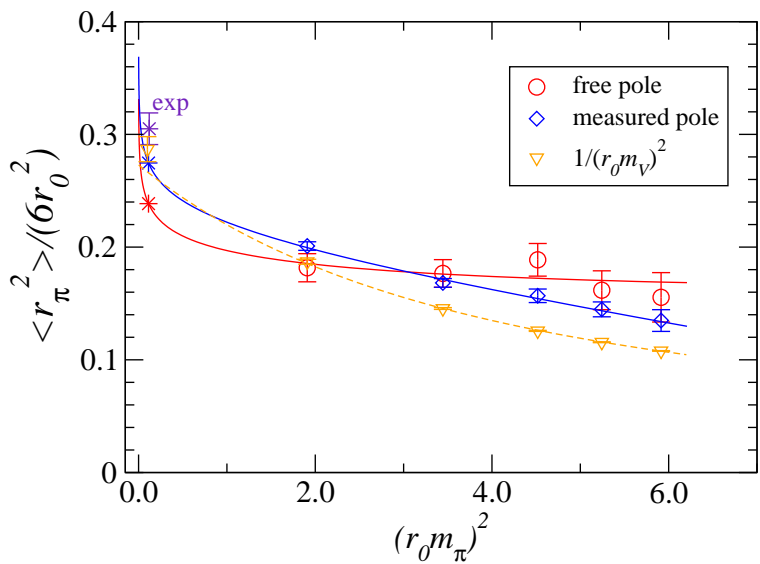}
  \caption{
    Chiral extrapolation of the pion charge radius.
    Results from the ``free pole'' (circles) and ``measured
    pole'' (diamonds) are plotted together with the 
    extrapolation curves with (\protect\ref{eq:chiral_fit}).
    An expectation from the pole dominance model $1/m_V^2$
    is also plotted (triangles).
  }
  \label{fig:chiral_extrap_knownpole}
\end{figure}

Chiral extrapolation of the pion charge radius is
plotted in Figure~\ref{fig:chiral_extrap_knownpole}.
The ``free pole'' fit results have essentially no dependence
on $m_\pi^2$ within the large statistical errors.
With the ``measured pole'' fit the statistical error is much
reduced and we find an upward trend toward the chiral limit.
It is strongly correlated with the pure pole dominance
model, which gives $6/m_V^2$ and the upward trend is
obvious.

For the chiral extrapolation we employ the one-loop ChPT
formula with a higher order analytic term
\begin{equation}
  \label{eq:chiral_fit}
  \langle r^2\rangle_V^\pi = C_0 
  - \frac{1}{(4\pi f)^2} \ln\frac{m_\pi^2}{\mu^2}
  + C_1 m_\pi^2,
\end{equation}
where $C_0$ and $C_1$ are fit parameters and $\mu$ is an
arbitrary scale.
We can see that the chiral logarithm enhances the charge
radius near the chiral limit, and the value extrapolated to
the physical point is
$\langle r^2\rangle_V^\pi$ = 0.396(10)~fm$^2$, which is
about 12\% below the experimental value.

\begin{figure}[tbp]
  \centering
  \includegraphics[width=10cm,clip]{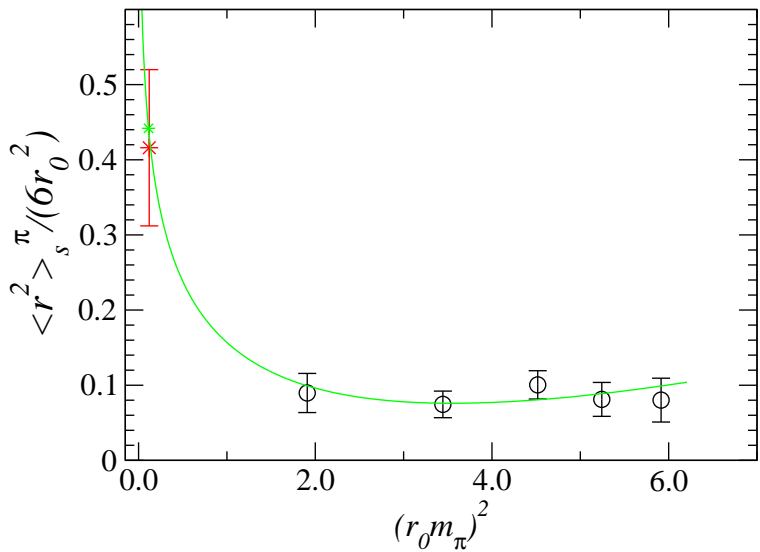}
  \caption{
    Chiral extrapolation of the scalar radius.
  }
  \label{fig:chiral_extrap_scalar}
\end{figure}

For the scalar form factor we do not expect the dominance of
the lowest-lying resonance (the $a_0$ meson), as it is
already rather far from the $q^2=0$ point.
We therefore simply fit the form factor by a polynomial.
Chiral extrapolation of the scalar radius is shown in
Figure~\ref{fig:chiral_extrap_scalar}, where we draw the fit
curve with the chiral log form plus a $O(m_\pi^2)$ analytic
term.
The chiral logarithm is very strong for this quantity and
the extrapolation could be sensitive to the details of the
fit function.
Our preliminary result neglecting such systematic effect is
0.60(15)~fm$^2$.

\bigskip
This work is supported by the Large Scale Simulation Program
No.~132 (FY2005) of High Energy Accelerator Research
Organization (KEK), and also in part by the Grant-in-Aid of
the Ministry of Education (Nos. 12740133, 13135204, 13640260,
14046202, 14740173, 15204015, 15540251, 16028201,
16540228, 16740147, 17340066, 17540259).

\end{document}